\documentclass[conference]{IEEEtran}
\IEEEoverridecommandlockouts
\usepackage{caption}
\usepackage{cite}

\usepackage{graphicx}%
\usepackage{multirow}%
\usepackage{amsmath,amssymb,amsfonts}%
\usepackage{amsthm}%
\usepackage{mathrsfs}%
\usepackage{xcolor}%
\usepackage{textcomp}%
\usepackage{manyfoot}%
\usepackage{booktabs}%
\usepackage{algorithm}%
\usepackage{algcompatible}%
\usepackage{algorithmicx}%
\usepackage{algpseudocode}%
\usepackage{listings}%
\usepackage{subfigure}

\theoremstyle{thmstyleone}%
\theoremstyle{thmstyletwo}%

\theoremstyle{thmstylethree}%
\newtheorem{definition}{Definition}%

\begin{document}

\title{Constrain Path Optimization on Time-Dependent Road Networks}

\author{\IEEEauthorblockN{Kousik Kumar Dutta}
\IEEEauthorblockA{
\textit{Indian Institute of Technology Ropar}\\
\textit{Department of CSE}\\
Rupnagar, India\\
kousik.21csz0004@iitrpr.ac.in}
\and
\IEEEauthorblockN{Venkata M. V. Gunturi}
\IEEEauthorblockA{
\textit{University of Hull}\\
\textit{School of Computer Science}\\
Hull, UK\\
v.gunturi@hull.ac.uk}
}

\maketitle

\begin{abstract}Time-Dependent Constrained Path Optimization (TD-CPO) takes the following input: (i) time-dependent (TD) road network, (ii) source ($s$), (iii) destination ($d$), (iv) departure time ($t$) and, (v) budget ($\mathcal{B}$). In TD graph, each edge is characterized by a time-dependent arrival time and a score function. TD-CPO aims to determine a loopless path $s$--$d$ departing from $s$ at time $t$ and arriving at $d$ on or before $t+\mathcal{B}$ while maximizing the score. TD-CPO has applications in urban navigation. TD-CPO is a variant of the Arc Orienteering Problem (AOP) known to be NP-hard in nature. The key computational challenge of TD-CPO is that we need to find the ``longest path'' in terms of score within the given budget constraint in a TD graph. Current works prune down the search space very aggressively. Thus, despite having low execution time, these algorithms often produce low-quality solutions. In contrast, our proposed approach $\mathcal{SCOPE}$ efficiently solves TD-CPO by exploiting road networks' spatial and temporal properties. The inherent computational structure of $\mathcal{SCOPE}$ enables trivial parallelization for improved performance. Our experiments indicate that $\mathcal{SCOPE}$ produces superior quality solutions (nearly $2x$) compared to the state-of-the-art algorithm while having comparable running times. Furthermore, $\mathcal{SCOPE}$ exhibits almost linear speedup as the number of CPUs (cores) increases (up to 24 CPUs).
\end{abstract}


\section{Introduction}\label{sec:intro}
Increasingly, the proliferation of mobility-based Big Data \cite{bigdata-shashi} has opened up new possibilities for more sophisticated routing queries on road networks (e.g., \cite{csp-foresthop,shahabi-ils,kkd-wise22,csp-td-vldb19,csp-td-icde19,shahabi-scenic}). For instance, one can now query paths like, ``Find a route for a 9:00 a.m. departure that traverses roads with the most scenic views \cite{shahabi-scenic}, with a constraint that the total length of the path does not exceed 15 minutes more than the fastest route'' or `Determine a path which maximizes safety (refer \cite{kaur-mdm}) for a 6:00 pm departure while constraining the total length of the path to be at most 10 minutes longer than the fastest path.''

The key aspect of both the previous queries is that they have a notion of maximizing a metric (scenic nature or safety) while constraining another metric (travel time). Both these queries can be formulated as variations of the classical Arc-Orienteering Problem (AOP) \cite{aop-GAVALAS2015}. It is important to note that the AOP is known to be NP-Hard \cite{aop-GAVALAS2015} in nature.

Authors in \cite{kkd-wise22,kaur-mnpj,kaur-mdm} have adapted the traditional arc orienteering problem definition for non-tourism-related path planning use cases. More specifically, they introduce a constraint of determining only loopless paths (i.e., nodes are not allowed to repeat) to make it more amenable for navigational scenarios on road networks. To this end, \cite{kkd-wise22,kaur-mnpj} presents the problem of Constrained Path Optimization (CPO), which aims to find a \textit{loopless} path between source and destination while \emph{maximizing a user preference metric} (e.g., safety, navigability, etc.) and adhering to certain constraints such as travel time, fuel, etc.

Current works in CPO (\cite{kaur-mdm,kaur-mnpj,kkd-wise22}) assume that the underlying road network is static in nature. In contrast, this work aims to develop a solution for the CPO problem on a road network, which is modelled as a time-dependent (TD) graph. Each edge in a TD graph is characterized by two parameters: a non-negative score and a travel time. These parameters are represented as time functions (details in Section \ref{sec:basic-concept}). Paths are typically interpreted as journeys through space and time in a TD graph. More specifically, in this setup, the travel time and score associated with each edge $e(u,v)$ in a candidate path $P$ is determined by the actual time at which a traveller would arrive at $u$ while following the designated path $P$. This kind of reference frame is widely used in the domain of routing \cite{tdcrp-icde19,csp-td-vldb19,shahabi-scenic} over road networks and has been formally referred to as a \emph{Lagrangian} reference frame by some authors \cite{GunturiS17,GunturiSY15}.

\noindent\textbf{Limitations of the prior work:} Research work closest to our problem lies in the area of engineered solutions for the Twofold Time-Dependent Arc Orienteering Problem (2TD-AOP) \cite{shahabi-scenic,memetic-j,memetic-conf}. 2TD-AOP finds a $s$--$d$ path between a given source $s$ and destination $d$, which maximizes the scenic view while constraining the travel time. The key difference between the 2TD-AOP path and the TD-CPO path is that 2TD-AOP allows the presence of a loop in the final path. This assumption is not always desirable in non-tourism-related scenarios, such as ours. The solution proposed in \cite{shahabi-scenic} is very conservative as it chooses the best arc to replace using the heuristic metric ``potential''. This aggressive pruning of the search space often exhibits lower solution quality. Moreover, adapting the 2TD-AOP solutions by removing loops in the final path typically results in low solution quality. The authors in \cite{memetic-conf,memetic-j} improve the solution in \cite{shahabi-scenic} by considering multiple initial seed paths instead of a single one (as \cite{shahabi-scenic}). Though the broader search space of the proposed algorithm produces better solution quality, this technique is not scalable for large datasets (around 10 seconds runtime for a graph with 6k nodes) (refer Section \ref{sec:exp}).

CSP\cite{csp-cola-gpu,csp-foresthop,csp-td-vldb19,csp-td-icde19,labeling} problem finds the shortest path between a source-destination pair subject to certain constraints. That means CSP minimizes the preference metric (distance) within some constraints, which differs from our TD-CPO problem, where one needs to maximize the preference metric within some constraints. Note that, it is not trivial to adapt the algorithms for minimization problems to solve maximization problems, as detailed next. 

\noindent \textbf{Challenges in adapting minimization-based approach:} A straightforward approach to reducing a maximization problem such as the CPO problem into a minimization problem (such as CSP) is to change the sign of score value (i.e., make it negative). However, minimization-based approaches (e.g., \cite{csp-cola-gpu,csp-foresthop,labeling}) inherently use a variant of Dijkstra's algorithm for optimization, which cannot work with negative score values. Another approach is to use reciprocals of the score values on edges, i.e., replace each score value $x$ with $\frac{1}{x}$, then use the minimization-based approach. However, with this reduction, we cannot get a one-to-one mapping with the original maximization-based CPO problem instance. The maximization problem involves maximizing the sum of scores of edges on the output path. Mathematically, it is equivalent to maximizing the sum of $n$ parameters, $s_1, s_2, \ldots s_n$ ($s_i$ denotes the score of edge $e_i$ in a path), which is not equivalent to minimizing the sum of their inverses $\frac{1}{s_1}$, $\frac{1}{s_2}$, $\ldots\frac{1}{s_n}$. To summarize, one cannot trivially generalize algorithms developed for minimization problems (e.g., \cite{csp-cola-gpu,csp-foresthop,labeling}) to obtain a solution for the maximization problem.

\noindent \textbf{Summary of Contributions:} This paper offers the following contributions:
    \begin{itemize}
       \item In this paper, we propose a novel algorithm called $\mathcal{SCOPE}$ to solve the CPO problem on a Time-Dependent graph. We call this \emph{TD-CPO problem}. 
        \item Our evaluation assesses the performance of $\mathcal{SCOPE}$ on large road networks (up to 0.68 million nodes and 1.8 million edges).
        \item In experiments, $\mathcal{SCOPE}$ produces good quality solutions within acceptable running times (within 3 secs) on large road networks. Our results indicate that \cite{memetic-j} could not scale beyond networks with 6k nodes, and \cite{shahabi-scenic} produces low-quality solutions on large networks. Furthermore, the linear scalability exhibited by $\mathcal{SCOPE}$ underscores its ability to scale up easily.
    \end{itemize}

\noindent\textbf{Outline:} The remainder of this paper is structured as follows. Section \ref{sec:basic-concept} discusses the fundamental concepts and formally defines the problem. After that, Section \ref{sec:background} presents some basic concepts used in our proposed algorithm. Section \ref{sec:proposed}  details our proposed solution, $\mathcal{SCOPE}$. Section \ref{sec:exp} presents experimental evaluation. Finally, we conclude in Section \ref{sec:conclusion}.

\begin{figure}[ht]
\centering
    \includegraphics[width=0.8\linewidth]{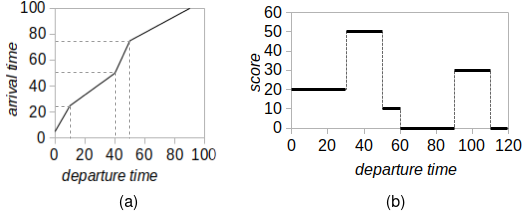}
\caption{(a) a continuous, non-decreasing ``piecewise-linear'' function as arrival time function, X and Y-axis is in time unit, (b) a ``piecewise-constant'' function as score function, X-axis is in time unit.}
\label{fig:function}
\end{figure}

\section{Problem Definition}\label{sec:basic-concept}
\begin{definition}[Road Network]
We represent the input road network using a temporally detailed graph (TD graph). TD graph represents road intersections as nodes and the road segments as directed edges. Each edge within the TD graph is associated with the following two functions. 

\noindent(a) Arrival Time Function ($\Gamma$): For any given departure time from the starting node, it computes the arrival time at the destination node along that edge.

\noindent(b) Score Function ($\Phi$): The score function determines the score or value associated with the edge for the given departure time from the starting node.

We choose time-dependent functions for both the arrival time and score function to achieve generalizability and better adaptability towards road network data in any format.

In our implementation, we have modelled the arrival time function as a continuous, non-decreasing ``piecewise-linear'' function (similar to current works \cite{csp-td-vldb19,csp-td-icde19} in the domain). At the same time, the score function is represented as a ``piecewise-constant'' function (similar to \cite{csp-td-vldb19,csp-td-icde19,shahabi-scenic,memetic-j}). It is important to note that this paper assumes First-In-First-Out (FIFO) property on the arrival time data on all edges. This implies that for any given edge $e(x,y)$, departing later from node $x$ will never arrive earlier at node $y$. Again, this assumption is in line with existing literature \cite{shahabi-scenic,csp-td-vldb19,csp-td-icde19}.

\noindent \textbf{Illustrating Arrival Time Function:} Consider an arbitrary edge $e(u,v)$ and timepoints ($x1, x2, \ldots$). Further assume that $e(u,v)$ exhibits travel times $c1, c2, \ldots$ and scores $s1, s2, \ldots$ for the departure-times $x1, x2, \ldots$. Thus, a journey departing from node $u$ at time $x1$ would arrive at node $v$ by time $y1 = x1 + c1$. Similarly, $y2 = x2 + c2$, etc. Note that, internally, we assume that travel time between two timepoints (say $x1$ and $x2$) will vary linearly i.e., increase (or decrease) from $c1$ to $c2$ (or remain the same if $c1=c2$). By combining timepoints and their corresponding arrival times, we create different breakpoints for edge $e(u,v)$, represented as ($x1, y1$), ($x2, y2$), $\ldots$. As mentioned earlier, we assume the function will vary linearly between two consecutive breakpoints. 
 
Fig. \ref{fig:function} (a) illustrates a sample arrival time function as a continuous non-decreasing ``piecewise-linear'' function. Given an arbitrary departure time $dt$ over an edge $e(u,v)$, we determine the arrival time at node $v$ is determined as follows. The arrival time function first identifies the closest timepoints (in the list of breakpoints of edge $e(u,v)$) between which the departure time $dt$ falls. Assume that $x1$ and $x2$ are the closest timepoints for $dt$. Thus, $x1 \leq dt \leq x2$, then the arrival time can be calculated as follows:

\begin{equation}
arrival\_time = (y2 - y1) \cdot \frac{dt - x1}{x2 - x1} + y1\label{eq:arr}
\end{equation}

Similarly, equation \ref{eq:dep} gives us the inverse function of equation \ref{eq:arr}. For a given arrival time at $v$ for edge $e(u,v)$, equation \ref{eq:dep} gives us the corresponding departure time from $u$.

\begin{equation}
    departure\_time = (x2-x1) \cdot \frac{arrival\_time-y1}{y2-y1} + x1\label{eq:dep}
\end{equation}

\noindent \textbf{Illustrating Score Function:} We model the score function as ``piecewise-constant'' function. Equation \ref{eq:score} represents a sample score function. 

\begin{equation}\label{eq:score}
score =
\begin{cases}
s1, & \text{if } x1 \leq departure\_time < x2 \\
s2, & \text{if } x2 \leq departure\_time < x3 \\
\vdots \\\\
\end{cases}
\end{equation}
\end{definition}


\subsection{\textbf{Time-Dependent Constrained Path Optimization (TD-CPO)}}\label{sec:problem}
\noindent \textit{Input:} Consists of the following: 

\noindent (1) A TD graph, $G=(V,E)$, where each directed edge $e(x,y)$ $\in$ $E$ is associated with an arrival time function ($\Gamma$) and score function ($\Phi$). 

\noindent (2) A source $s$ $\in$ $V$ and a destination $d$ $\in$ $V$.

\noindent (3) Departure time $t_{dep}$ from $s$.

\noindent (4) A positive value $overhead$ corresponds to the maximum permissible travel time allowed over the travel time of the fastest path from $s$ to $d$. We use the term ``budget ($\mathcal{B}$)'' to denote the sum of overhead and the travel time of the fastest path from $s$ to $d$. 

\noindent \textit{Output:} A loopless path $\mathcal{P}^*$ between $s$ and $d$. 

\noindent \textit{Objective function:} Maximize $\Phi(\mathcal{P}^*)$ 

\noindent \textit{Constraint:}  $\Gamma(\mathcal{P}^*) \leq t_{dep} + \mathcal{B}$

\subsection{Generalization of TD-CPO to Multiple Constrains}
TD-CPO can be trivially generalized to consider multiple constraints. In a multi-constrained version of the TD-CPO problem, the input and objective functions will be the same as those of TD-CPO, with the only changes being in the constraints. We can include multiple constraints by defining a budget for each of the different constraint metrics, e.g., $\mathcal{B}1$, $\mathcal{B}2\ldots$. Now, the goal would be to find path $\mathcal{P}^*$ such that $\Gamma1(\mathcal{P}^*) \leq \mathcal{B}1;$ $\Gamma2(\mathcal{P}^*) \leq \mathcal{B}2\ldots$. Here, $\Gamma1, \Gamma2, \ldots$ are functions to measure the path $\mathcal{P}^*$ according to different constraints. 

Section \ref{sec:generelize} details the minimal changes that need to be done in our proposed algorithm to consider multiple constraints. Note that the Arc-Orienteering Problem (AOP) was formally defined to contain only a single constraint. We intentionally chose a different terminology (TD-CPO) to distinguish our work from AOP on the lines of \emph{loopless paths} and \emph{generalization to multiple constraints}.

\section{Basic Concepts for the Proposed Algorithm}
\label{sec:background}
\subsection{\textbf{MINSUM Algorithm}}
\label{sec:labeling}
In this section, we discuss the MINSUM algorithm \cite{labeling}. MINSUM aims to determine a path between a source and a destination that minimizes the distance subject to a constraint on travel time. Its exploration strategy is similar to Dijkstra's and involves labelling nodes and relaxing edges (and updating labels). The predecessor information is also stored for each label to retrieve the computed path. However, unlike Dijkstra's, it needs to maintain two parameters (distance and travel time in \cite{labeling}) in a label. Additionally, it does not use Dijkstra's concept of closing a node (corresponding to the result of extract-min). Instead, it discards unnecessary labels using the concept of dominated labels as defined below. Note that given the nature of dominated labels, each node may be a collection of non-dominated labels as the algorithm progresses. 

\begin{definition}[Dominated label]\label{def:domination}
Suppose $L1(t1,s1)$ and $L2(t2,s2)$ are two different labels of a node $v$. Now $L1$ is dominated by $L2$ if: (a) $t1>t2$, and (b) $s1>s2$.
\end{definition}

\begin{figure*}[ht]
\centering
    \includegraphics[width=0.7\linewidth]{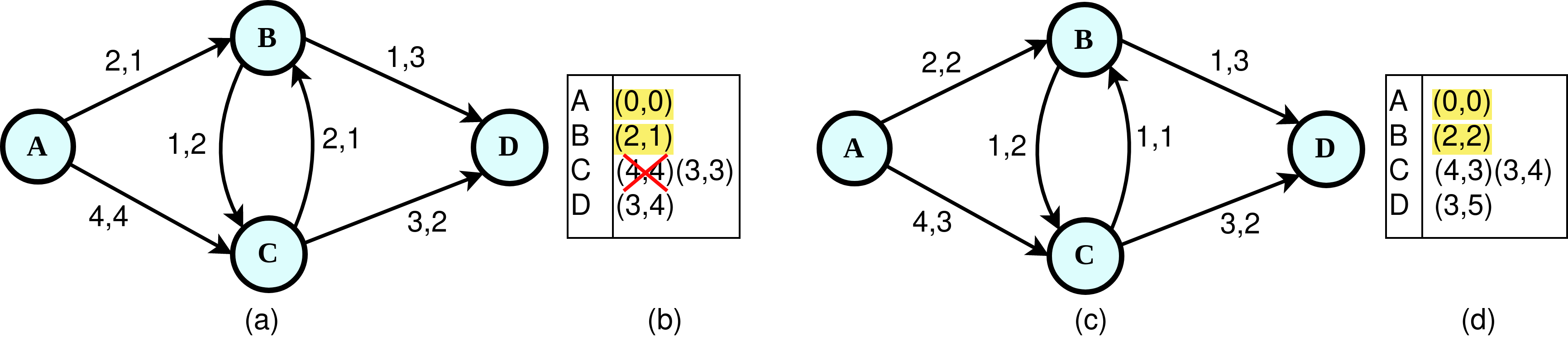}
\caption{(a) \& (b) sample graph and dominated label pruning for shortest path problem, (c) \& (d) sample graph and example to show that the dominated label can not be pruned for loopless longest path problem.}
\label{fig:domination}
\end{figure*}

We will now discuss the working of the algorithm procession for source $A$ and destination $D$ using the graph illustrated in Fig. \ref{fig:domination}(a). Each edge is characterized by two parameters: $(l, c)$ represents distance and travel time, respectively. Initially, the algorithm sets up a min-heap to store labels in ascending order of travel time and maintains a node-wise list to keep track of all non-dominated labels associated with each node (refer Fig. \ref{fig:domination}(b)).

The algorithm first labels the source node ($A$) with a label of (0,0). After each expansion, the algorithm inserts a new label into the min-heap if and only if (a) the new label is not dominated by any existing label (based on Definition \ref{def:domination}) and (b) the associated travel time is less than the given budget constraint. In the example shown in Fig. \ref{fig:domination}(b), node $C$ is labelled with $L1$(4,4) and $L2$(3,3), where $L1$ is dominated by $L2$, leading the algorithm to discard $L1$ from further processing. The algorithm continues until the min-heap becomes empty. All Pareto optimal paths from node $A$ to node $D$ are stored in the label list of node $D$. The optimal path is then selected from these Pareto optimal paths as the one with the minimum travel time.

\subsection{\textbf{Inspiration for exploration strategy for TD-CPO}}
\label{sec:challenge}
MINSUM algorithm minimizes the distance within a travel time constraint, whereas the TD-CPO problem is framed as a maximizing score within the same budget constraint. Hence, the TD-CPO problem becomes identifying the longest path in terms of score while adhering to budget constraints. However, determining the longest path in a cyclic graph is an NP-hard problem, and it can only be solved in linear time in a directed acyclic graph (DAG) (\cite{longestpath}). Converting a cyclic graph into a DAG is non-trivial (\cite{cyclic-to-dag}). 

In addition, we cannot use the concept of dominated label pruning for our problem definition (i.e., find a loopless path while maximizing the score within the budget constraint). Consider the example shown in Fig. \ref{fig:domination}(d); there are two labels (3,4) and (4,3) at node $C$. However, label (3,4) cannot visit node $B$ again due to loop constraints as discussed in TD-CPO problem definition (Section \ref{sec:problem}). Hence, leading to label (6,6) at node $D$. Meanwhile, label (4,3) can visit node $B$ and produce label (6,7) at node $D$, achieving a higher score. Thus, finding the longest path with loopless constraints complicates label pruning based on domination.

In summary, the TD-CPO problem's maximization goal, the complexity of finding the longest path in cyclic graphs, the challenge of converting cyclic graphs to DAGs, and the difficulty of adapting dominant label pruning strategies for loopless paths create significant computational challenges. These challenges require the development of specialized algorithms to solve TD-CPO instances. However, we are inspired by the fundamental exploration strategy of MINSUM \cite{labeling} and use some of the concepts while designing our $\mathcal{SCOPE}$ algorithm to solve the TD-CPO problem.

\begin{figure*}[ht]
\centering
    \includegraphics[width=0.9\linewidth]{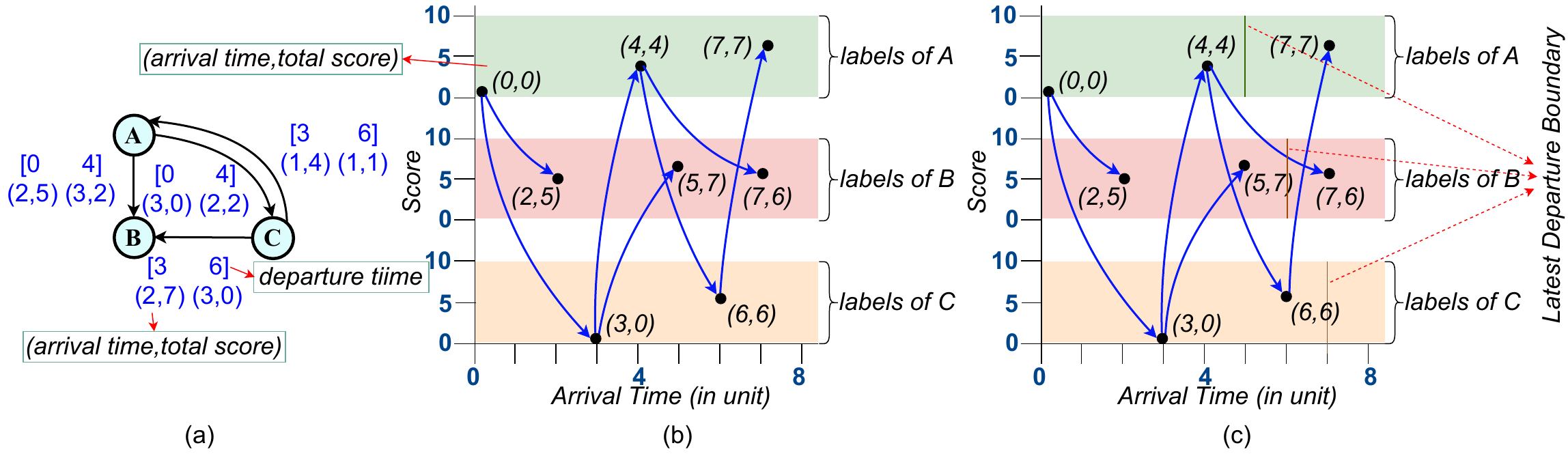}
\caption{(a) A sample TD graph, (b) Labels created by the naive algorithm for departure time 0 and budget 8, (c) Temporal pruning using Latest Departure Boundary.}
\label{fig:LP-graph}
\end{figure*}

\subsection{Temporal Pruning Strategy}
\label{sec:prunning}
As mentioned earlier, we cannot use the concept of dominated label pruning for TD-CPO. To this end, we use temporal pruning to reduce the number of labels needed to be processed for a solution. Here, we use the concept of the \textit{latest departure time} as used in the latest departure path problem in \cite{GunturiS17}.

\begin{definition}[Latest departure time]
The latest departure time of a node is defined as the latest time to depart from that node to reach the destination by the desired arrival time, i.e., departure time from source + budget.
\end{definition}

The latest departure path problem is solved by the \emph{Backward Traversal} algorithm. 
\emph{Backward Traversal} is similar to the latest departure path problem \cite{GunturiS17,aop-GAVALAS2015}. It determines the latest departure time from a node $x$ to reach the destination $d$ by desired arrival time $t_{arr} = t_{dep} + \mathcal{B}$ (for departure time from source node $s$ as $t_{dep}$ and budget $\mathcal{B}$). It follows Dijkstra's styled enumeration. However, it traverses in the reverse direction from the destination node and relaxes the incoming edges. The priority queue is a max heap (instead of a min heap). In each iteration, we close the node with the maximum label. The B-labeling algorithm terminates when (a) the heap is empty or (b) the top of the heap has a label that is less than $t_{dep}$.

The backward traversal algorithm labels all nodes with their latest departure times (to reach the destination by the desired arrival time). Therefore, any label (temporal copy) of a node with an arrival time (as determined by the naive algorithm) greater than the latest departure time for that node cannot contribute to the valid TD-CPO path in the future and hence can be pruned. Fig. \ref{fig:LP-graph}(c) illustrates an example of temporal pruning, where the latest departure times for nodes $A$, $B$, and $C$ are 5, 6, and 7, respectively. Temporal pruning is done through the concept of what we refer to as the \emph{Latest Departure Boundary}.

\begin{definition}[Latest Departure Boundary]
    Given an instance of the TD-CPO query, for any node $v$ in the input TD graph $G(V,E)$, the \emph{Latest Departure Boundary} is defined as the latest departure time from $v$ to reach the given destination $d$ by the desired arrival time of $t_{dep} + \mathcal{B}$. $t_{dep}$ is the departure time from the corresponding source $s$ and $\mathcal{B}$ is the maximum permissible budget.
\end{definition}

The latest departure boundary for nodes $A$, $B$, and $C$ are illustrated in Fig \ref{fig:LP-graph}(c). The latest departure boundaries for nodes $A$, $B$, and $C$ are $t=5$, $t=6$ and $t=7$ respectively. These boundaries define the ``furthest temporal copy'' (or labels) of the nodes, which can result in a feasible path of the given TD-CPO query. For example, the label (7,7) of node $A$ is infeasible (and hence pruned) as its label $7$ is outside the latest departure boundary $t=5$ for node $A$.

\section{Proposed Algorithm}
\label{sec:proposed}
 In this section, we first discuss the working of $\mathcal{SCOPE}$. Finally, We discuss the techniques to parallelize $\mathcal{SCOPE}$ using multiple CPU cores.

\subsection{\textbf{Details of $\mathcal{SCOPE}$} Algorithm}\label{sec:scope}
We first invoke the \emph{backward traversal} algorithm \cite{GunturiS17} from destination $d$ to mark all the reachable nodes $v$ with the latest departure time ($\mathcal{K}_{t_{arr}}(v)$) such that one can reach the destination by the desired arrival time of $t_{arr}$ (departure time from source $t_{dep}$ + budget $\beta$). If the backward traversal algorithm does not mark a node with a label greater than (or equal to) $t_{dep}$, it cannot be a part of the solution. This implies that if the backward traversal algorithm marks the source node $S$ with a label less than $t_{dep}$, there is no feasible solution for the given TD-CPO query instance.           

\begin{algorithm}[!ht]
\caption{$\mathcal{SCOPE}$}\label{Alg:scope}
\begin{flushleft}
\textbf{Input:} (a) Input TD graph $G(V,E)$; (b) source node $s$; (c) destination node $d$; (d) departure time from $s$: $t_{dep}$; (e) budget $\Theta$.\\
\textbf{Output:} The optimal path $\mathcal{P^*}$, which start from $s$ at $t_{dep}$, and reach to $d$ by $t_{arr} \Leftarrow t_{dep} + \Theta$.
\end{flushleft}
\begin{algorithmic}[1]
\State $\mathcal{K}_{t_{arr}} \leftarrow$ Determine the latest departure time of all the nodes (using the backward traversal algorithm) to able to reach $d$ by $t_{arr} = t_{dep} + \Theta$. \State //$\mathcal{K}_{t_{arr}}(v)$ is the latest departure boundary of $v$. It should be at least $t_{dep}$ for $v$, otherwise $v$ cannot be a part of any solution.
\If{label of $s$ $<$ $t_{dep}$}\\
    \Return  //No feasible solution possible for the given TD-CPO query
\EndIf
\State Create label for the source node $s$ $<s,t_{dep},0,null>$ 
\State Create a visited list $vl$ corresponding to the label $<s,t_{dep},0,null>$
\State Add node $s$ to the visited list $vl$ of the label $<s,t_{dep},0,null>$ 
\State $optimalLabel \Leftarrow$ ProcessLabel($<s,t_{dep},0,null>$, visited list $vl$)
\State $\mathcal{P^*} \Leftarrow optimalLabel.getPath()$ //retrieves the optimal path from the predecessor list of the $optimalLabel$.\\
\Return $\mathcal{P^*}$

\end{algorithmic}
\end{algorithm}

\begin{algorithm}[!ht]
\caption{ProcessLabel($<u,\alpha,sc,pred>$, $vl'$)}\label{func:processlabel}

\begin{algorithmic}[1]
\If{$u$ == $d$} \label{line:ln2}\\
    \Return $<u,\alpha,sc,pred>$ \label{line:ln3}
\EndIf
\State $resultLabel \leftarrow NULL$
\ForAll{adjacent node $v$ of $u$}   \label{line:ln6}
    \If{$v$ was never visited in $vl'$} \label{line:ln8}
        \State $v_{at} \leftarrow u.\Gamma(v)_{\alpha}$\label{line:comp}
        \State //Process the new label, if $v_{at}$ less than latest departure boundary of the node $v$.\label{line:com1}
        \If{$v_{at}\leq \mathcal{K}_{t_{arr}}(v)$}\label{line:cond1}
            \State $v_{sc} \leftarrow sc +  u.\Phi(v)_{\alpha}$
            \State $v_{pred} \leftarrow$ $<u,\alpha,sc,pred>$
            \State Create the new label $<v,v_{at},v_{sc},v_{pred}>$
            \State $vl''$ $\leftarrow$ Copy of $vl'$ \label{line:ln14}  
            \State Add $v$ to the visited list $vl''$
            \State $label \Leftarrow$ ProcessLabel($<v,v_{at},v_{sc},v_{pred}>$, $vl''$)
            \If{$label \neq NULL$}
                \State Check and update the $resultLabel$ with $label$ if $label$ has better score.\label{line:ln19}
                
            \EndIf
        \EndIf
    \EndIf
\EndFor\\
\Return $resultLabel$.
\end{algorithmic}
\end{algorithm}

Following this, we start the forward search of the algorithm. We initialize the source node $s$ with a label. In our algorithm, a label is a tuple that consists of the following information: (i) physical node, (ii) arrival time at the node, (iii) total accumulated score along the path, and (iv) predecessor label. For example., a label $<s,t_{dep},0,null>$ means the following: the path (from source) through the designated predecessor ($null$ in this case) arrives at $s$ at time $t_{dep}$ and has a current total accumulated score of $0$. Here, each unique label is linked to a visited list, which lists all the physical nodes that will be visited if one wants to follow the predecessors of the labels until one reaches the source node. This visited list is used to prevent the creation of loops in the candidate paths under consideration. Note that, unlike greedy algorithms such as the Dijkstra's, we cannot close any nodes. Thus, we have multiple visited lists, one for each candidate path the algorithm explores. Following the creation of the label, we create a visited list $vl$ for the label $<s,t_{dep},0,null>$ and add $s$ to $vl$. 

\begin{figure*}[!ht]
    \centering
    \includegraphics[width=0.9\linewidth]{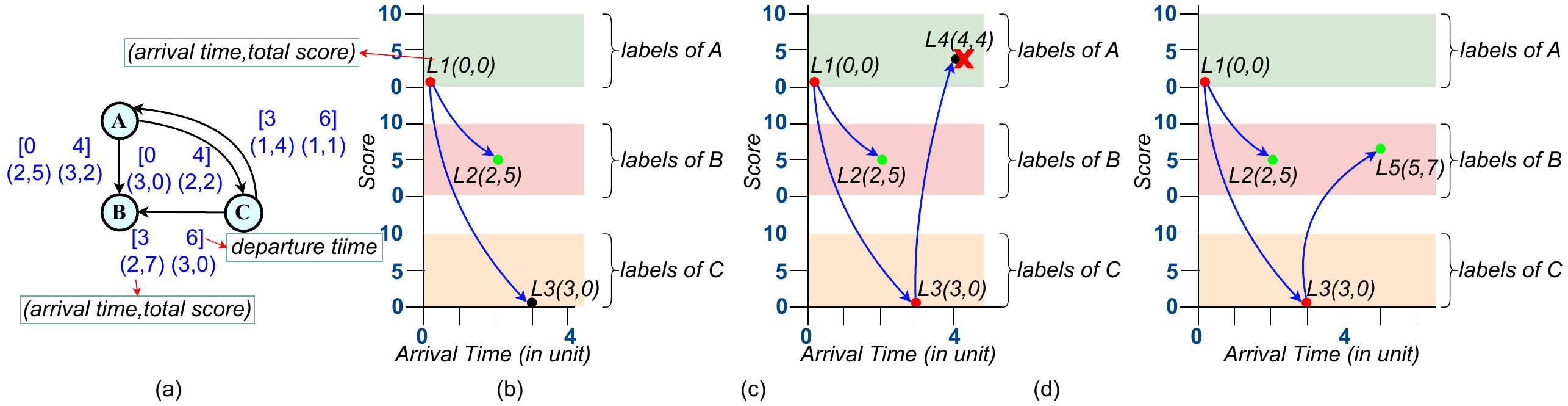}
    \caption{(a) sample TD graph, (b), (c), (d) a sample TD-CPO query procession for source $A$, destination $B$, departure time from $A$ as $t=0$, budget $8$.}
    \label{fig:lp-procession}
\end{figure*}

After initialization, the algorithm calls Algorithm \ref{func:processlabel} with label $<s,t_{dep},0,null>$ and visited list $vl$. Algorithm \ref{func:processlabel} proceeds recursively to find the $resultLabel$ for the destination node $d$ to maximize the score while being within the budget. At the termination of the recursion, the resulting path is obtained using the predecessor information. We now detail the recursive procedure.

Consider an arbitrary $i^{th}$ recursion call of the ``ProcessLabel'' procedure (Algorithm \ref{func:processlabel}). Assume it was invoked with label $<u,\alpha,sc,pred>$ and visited list $vl'$. The procedure then expands the label $<u,\alpha,sc,pred>$ recursively for each adjacent node $v$ of $u$. Firstly, it checks if $v$ is in the visited list $vl'$ or not (line \ref{line:ln8} in Algorithm \ref{func:processlabel}). If $v$ is already in $vl'$, adding it again creates a loop in the solution. Thus, we process $v$ only if it is not in the  $vl'$.

Afterwards, it computes the arrival time at $v$ ($v_{at}$) along the edge $u$-$v$ for a departure time $t=\alpha$ at $u$. Temporal pruning is then applied by verifying whether $v_{at}$ is less than or equal to the latest departure boundary from $v$ ($\mathcal{K}_{t_{arr}}(v)$). Subsequently, the total accumulated score at $v$ ($v_{sc}$) is computed by adding the total accumulated score at $u$ to the score gain from travelling along edge $u$-$v$ at time $\alpha$. The label $<u,\alpha,sc,pred>$ is set as the predecessor of $v$ ($v_{pred}$), and a new label $<v,v_{at},v_{sc},v_{pred}>$ is created for further recursion. 

To construct visited list $vl''$ for further recursion, we check if $u$ has several adjacent neighbours $v1, v2\ldots$. A new visited list is created for all the neighbours by coping the $vl'$, except for one. For one of the neighbours, $vl''$ is essentially a reference $vl'$ itself. And then, we add $v$ to the $vl''$. Finally we call the ``ProcessLabel'' procedure recursively with parameters $<v,v_{at},v_{sc},v_{pred}>$ and $vl''$.

Suppose the label returned by the recursion call is not $NULL$ and has a score better than the current $resultLabel$. In that case, we update the $resultLabel$ with the returned label (line \ref{line:ln19} in Algorithm \ref{func:processlabel}). The recursion call terminates if the current label is a temporal copy of the destination $d$ (lines \ref{line:ln2}--\ref{line:ln3} in Algorithm \ref{func:processlabel}). This implies that we have found a solution for the given TD-CPO query with source $s$, destination $d$, budget constraint $\mathcal{B}$ for the departure time $t_{dep}$ at $s$. Further explanation on the processing of a single recursion call of $\mathcal{SCOPE}$ is provided through a sample example in Fig. \ref{fig:lp-procession}.

In Fig. \ref{fig:lp-procession}, we illustrate the execution of the $\mathcal{SCOPE}$ algorithm. Fig. \ref{fig:lp-procession}(a) depicts the example graph with nodes $A$, $B$, $C$, and their connecting edges, each annotated with respective time series data representing edge properties. Here, we represent the edge properties as time series data instead of continuous functions for easier understanding. Fig. \ref{fig:lp-procession}(b), \ref{fig:lp-procession}(c), \ref{fig:lp-procession}(d), and \ref{fig:lp-procession}(e) demonstrate the algorithm's progression for a sample query with source $A$, destination $B$, departure time $0$, and budget $8$. $\mathcal{SCOPE}$ initiates execution by initializing the source label $L1$ with $<A,0,0,null>$. However, for easier understanding, we only present the arrival time and total accumulated score up to that point in a label representation in Fig. \ref{fig:lp-procession}. Moreover, we omit the other two parameters as the physical node component is represented by different colour blocks, and the predecessor of a label can be tracked by the edges in Fig. \ref{fig:lp-procession}.

After initialization, $\mathcal{SCOPE}$ calls the ``ProcessLabel'' function to recursively expand $L1$. During this process, ``ProcessLabel'' generates $L2$ ($<B,2,5,L1>$) and $L3$ ($<C,3,0,L1>$) from $L1$. Subsequently, it recursively calls itself with input label $L3$ (as $L2$ is a temporal copy of destination $B$). Upon processing $L3$, it creates $L4$ ($<A,4,4,L3>$). However, $L4$ is discarded from the further recursion as it is another temporal copy of $A$ along $L1$ in the path $L1$--$L3$--$L4$. $L3$ also generates label $L5$ ($<B,5,7,L3>$) in the next step. At this point, $L1$ and $L2$ are already processed, and $L3$ and $L5$ are temporal copies of destination $B$. Therefore, the algorithm terminates and returns $L5$ as the $resultLabel$ since it has a higher score than $L2$.

\noindent\textbf{Multi-thread Execution of $\mathcal{SCOPE}$}
\label{sec:multicore}
Due to the recursive nature of $\mathcal{SCOPE}$, developing a parallel implementation with good CPU utilization is not trivial. In \cite{arxiv}, the authors discuss the limitations of trivial parallelization in a recursive algorithm for a related problem. The ``ProcessLabel'' function of $\mathcal{SCOPE}$  involves numerous recursive calls for each adjacent node of the $currentLabel$. We have implemented the parallel approach utilizing the ``work-stealing'' scheduler provided by the ``Java ForkJoinPool'' library. The ``work-stealing'' scheduler is well-suited for parallelization of tasks involving multiple levels of recursive calls. \textbf{By executing these recursion calls in parallel, we achieve a linear scale-up in performance as the number of cores increases.}

\noindent\textbf{Generalization to Multi Constrained TD-CPO}\label{sec:generelize}
As mentioned earlier, our proposed algorithm can be trivially generalized to consider multiple constraints. We now briefly describe this generalization. In Algorithm \ref{func:processlabel}, we check whether the new label adheres to the given constraint or not (here it is travel time) in line \ref{line:com1}, \ref{line:cond1}. We can easily add different checking conditions here for different constraints to generalize $\mathcal{SCOPE}$ for multiple constraints. As the complexity of checking whether the new node matches a single constraint is $O(1)$, adding multiple constraints would also not affect the runtime of $\mathcal{SCOPE}$.

\noindent\textbf{Time Complexity Analysis}\label{sec:omplexity}
$\mathcal{SCOPE}$ initially sets the source label and invokes the ``ProcessLabel'' function. After that, the ``ProcessLabel'' function recursively calls itself for all the adjacent nodes of the $currentLabel$ until it reaches the destination or exceeds the budget. Considering that the TD-CPO problem definition prohibits loops in the final solution, in the worst case, there may be a total of $|V|$ nodes (for graph $G(V,E)$) in the final path. If we consider the maximum degree of the graph as $\Delta$, the worst-case time complexity would be $O(\Delta^{|V|-1})$. \textbf{Despite having exponential time complexity, $\mathcal{SCOPE}$ achieves efficient performance in real-world road networks due to their lower degree of connectivity and temporal characteristics.}

\begin{table}[ht]
\centering
\caption{DATASETS}\label{tab::dataset}%
\begin{tabular}{@{}l|ccc@{}}
\toprule
\textbf{Road Network } & \textbf{\#Nodes } & \textbf{\#Edges }  & \textbf{Average Edge Length }\\
			\midrule
            \textbf{Oldenburg} & 6105 & 7035 & 74 meters \\
			\textbf{Bengaluru} & 149028 & 396726 & 123 meters  \\
			\textbf{NewYork} & 334930 & 914391 & 179 meters \\
			\textbf{Moscow} & 685091 & 1868268 & 	133 meters\\
\hline
\end{tabular}
\end{table}
\begin{table}[ht]
\centering
\caption{\textbf{Rush hour timings in the different cities}}\label{tab::rush-time}%
\begin{tabular}{@{}l|c|c@{}}
\toprule
\textbf{Road Network } & \textbf{Morning } & \textbf{Evening}  \\
			\midrule
   \textbf{Oldenburg} & 8:00a.m.-11:30a.m. & 5.30p.m.-8:00p.m.   \\
			\textbf{Bengaluru} & 7:00a.m.-11:00a.m. & 5.00p.m.-8:00p.m.   \\
		   \textbf{NewYork} & 7:30a.m.-9:30a.m. & 5.00p.m.-7:00p.m.  \\ 
     \textbf{Moscow} & 7:00a.m.-10:00a.m. & 6.00p.m.-9:00p.m.   \\
\hline
\end{tabular}
\end{table}

\section{Experimental Evaluation}
\label{sec:exp}
\noindent \textbf{Dataset} Our experimental analysis focuses on four road networks. Bengaluru, New York, and Moscow sourced from \cite{datacite}. Relatively smaller Oldenburg network from \cite{datacite2}. Table \ref{tab::dataset} provides more details of these datasets. For our experiments, we synthetically generate travel time data (travel time of an edge and score data) on these road networks.

To account for time-dependent travel times of an edge, we simulate rush hour congestion patterns, considering two daily rush hours: one in the morning and the other in the evening. Table \ref{tab::rush-time} presents the details of the duration of the rush hours on different road networks. During these defined rush hours, we gradually increase travel times, starting from the beginning of each city's rush hour and peaking at the midpoint of the rush hour duration. The extent of the travel time increase at its peak varies randomly between $30\%$ and $35\%$. After reaching the peak, travel times are progressively reduced until they return to their original values at the end of the rush hour. We record these changes at half-hour intervals during the rush hour, serving as distinct timepoints in the arrival time and score functions.

These time-dependent variations are uniformly applied to all edges during rush hours. Non-rush hour travel times for each edge are computed by dividing the edge's distance by a random speed value within the allowable speed limits in the city, ranging from the maximum ($400 meters/minute$) to the minimum ($250 meters/minute$), resulting in travel times represented in minutes. 

Furthermore, we evenly distribute edges with positive scores across the network, with the proportion of such edges varying between $10\%$, $20\%$, and $30\%$ of the total edges. For each selected edge, we assign a random integer between 0--15 to the score function. The remaining edges receive a score of 0.

\noindent \textbf{Query set formation}\label{sec:query-set} 
We generate four query sets per dataset with varying budget ranges. Each set includes 200 randomly chosen source-destination pairs. The process involves selecting a source, a destination, and a departure time within a rush hour, as the travel time of an edge or score is static outside the rush hour. After that, we calculate the total budget with a fixed parameter called ``overhead''. These pairs are grouped into query sets based on their budget ranges. Set-1 covers pairs with budgets of $0$--$5$ minutes, set-2, set-3 and set-4 spans $5$--$10$ minutes, $10$--$15$ minutes and $15$--$20$ minutes, respectively. For each set, we report the average runtime (in seconds) and the average score of the output paths within that budget range.

\subsubsection{Candidate algorithms}
    \noindent\textit{REC-INSERT \cite{shahabi-scenic}}, \textit{(b) MEMETIC \cite{memetic-j}:} We compare $\mathcal{SCOPE}$ with the state-of-the-art 2TD-AOP solvers, adapted for TD-CPO, by eliminating loops (if any) from the final path.

\subsubsection{Experimental setup} We implemented all the algorithms in Java and utilized OpenJDK-17 to build and run all the algorithms. Our experiments were conducted on the Intel Xeon Gold 6258R server with 28 cores (56 threads) and 32GB of RAM. The processor operates at a base frequency of 2.7GHz, with a maximum boost frequency of 4GHz.

\begin{figure}[t]
\centering
    \subfigure[Score comparison in Oldenburg network for 50\% overhead, 20\% edges with positive score.]{\includegraphics[width=0.45\linewidth]{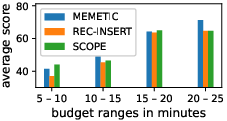}\label{fig:OL_score}}
    \quad
\subfigure[Runtime comparison in Oldenburg network for 50\% overhead 20\% edges with positive score. Y-axis is in $log_{100}$ scale.]{    \includegraphics[width=0.48\linewidth]{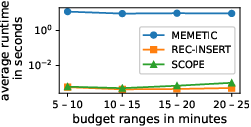}\label{fig:OL_time}}
\caption{Comparison between MEMETIC, REC-INSERT and $\mathcal{SCOPE}$.}
\label{fig:smaller-exp}
\end{figure}

\begin{figure}[t]
\centering
    \subfigure[Score comparison in Bengaluru network for 30\% overhead, 20\% edges with positive score.]{
    \includegraphics[width=0.45\linewidth]{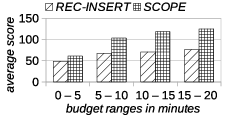}\label{fig:bengaluru_score}}
    \quad
    \subfigure[Score comparison in New York network for 30\% overhead, 20\% edges with positive score.]{
    \includegraphics[width=0.45\linewidth]{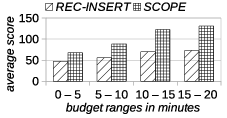}\label{fig:newyork_score}}
    \subfigure[Runtime comparison with varying number of cores in Bengaluru network for 30\% overhead, 20\% edges with positive score. Y-axis is in $log_{100}$ scale.]{
    \includegraphics[width=0.45\linewidth]{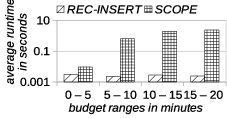}\label{fig:bengaluru_time}}
    \quad
    \subfigure[Runtime comparison with varying number of cores in New York network for 30\% overhead, 20\% edges with positive score. Y-axis is in $log_{100}$ scale.]{
    \includegraphics[width=0.45\linewidth]{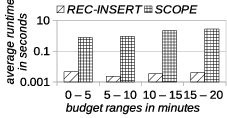}\label{fig:newyork_time}}
\caption{Comparison between REC-INSERT and $\mathcal{SCOPE}$ for Bengaaluru and New York road networks.}
\label{fig:rec-insert}
\end{figure}

\begin{figure}[t]
\centering
    \subfigure[Score comparison in Moscow network for 30\% overhead, 20\% edges with positive score.]{\includegraphics[width=0.47\linewidth]{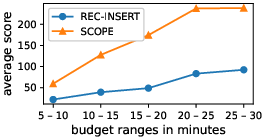}\label{fig:moscow_score}}
    \quad
    \subfigure[Runtime comparison with varying number of cores in Moscow network for 30\% overhead, 20\% edges with positive score.]{\includegraphics[width=0.45\linewidth]{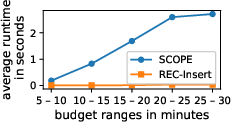}\label{figmoscow_time}}
\caption{Comparison between REC-INSERT and $\mathcal{SCOPE}$ in Moscow network.}
\label{fig:moscow}
\end{figure}

\subsection{\textbf{Experimental Evaluation}}
\label{sec:results}
\subsubsection{Performance Evaluation}
\label{sec:performance}
This section presents the performance of $\mathcal{SCOPE}$. Firstly, we conduct a comparative analysis of $\mathcal{SCOPE}$ against MEMETIC and REC-INSERT algorithm for the smaller Oldenburg network. There after, we compare all the three algorithms in the larger Bengaluru, New York, and Moscow network. Following that, we assess the runtime of $\mathcal{SCOPE}$ while varying the number of cores. We have maintained a fixed overhead of $30\%$ and a $20\%$ number of edges with a positive score value for these experiments. The experiments are conducted using 24 cores.

Fig. \ref{fig:OL_score} and \ref{fig:OL_time} illustrate the performance of $\mathcal{SCOPE}$ in smaller Oldenburg network. MEMETIC has slight better score, while REC-INSERT and $\mathcal{SCOPE}$ have comparable score (Fig. \ref{fig:OL_score}). In terms of runtime, REC-INSERT and $\mathcal{SCOPE}$ has similar runtime of \texttt{1 ms}, while MEMETIC has exponential runtime over \texttt{10 secs}. To this end, the smaller network doesn't test REC-INSERT and $\mathcal{SCOPE}$ as both perform similarly. MEMETIC doesn't show scalability, hence doesn't considered for the further experiments.

Fig. \ref{fig:rec-insert} illustrates the performance of $\mathcal{SCOPE}$ compared to REC-INSERT across the road networks. $\mathcal{SCOPE}$ achieves a superior score gain compared to the heuristic REC-INSERT. $\mathcal{SCOPE}$ attains a maximum score gain of $3$ times for  the Moscow dataset and a minimum score gain of $1.24$ times for query the Bangalore dataset, as compared to the average score gain achieved by REC-INSERT.

Regarding runtime, REC-INSERT outperforms $\mathcal{SCOPE}$ for all query sets. Nonetheless, $\mathcal{SCOPE}$ maintains a reasonable runtime. For query set-4 from the New York dataset, $\mathcal{SCOPE}$ records a maximum average runtime of $2.74$ seconds. Given the aggressive nature of REC-INSERT, it consistently completes execution within one second. In contrast, $\mathcal{SCOPE}$ explores larger search space to return better solution. Therefore, while $\mathcal{SCOPE}$ exhibits longer runtime, its applicability in cases where high-quality solutions and feasible runtime needs provide an advantage.

\begin{figure}[ht]
\centering
    \includegraphics[width=0.55\linewidth]{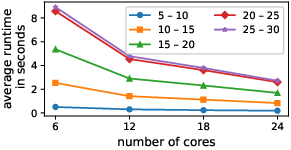}\label{}
\caption{Linear scale up showed by the multi-core execution of $\mathcal{SCOPE}$ for different budget ranges ($5$--$10$, $10$--$15$, $15$--$20$, $20$--$25$ , and $25$--$30$ minutes) in Moscow network for overhead of $30\%$ over the fastest path cost and $20\%$ of edges with positive score.}
\label{fig:scale-up}
\end{figure}

Fig. \ref{fig:scale-up} presents the results of $\mathcal{SCOPE}$ executed with different core counts and assesses the linear scale-up achieved by $\mathcal{SCOPE}$ with increasing numbers of cores (6, 12, 18, 24). For query set-1, the execution time is already within one second. Thus, it cannot scale further despite an increasing core count. For query set-2, set-3, set-4 and set-5, $\mathcal{SCOPE}$ demonstrates nearly linear scalability, and the curve saturates as the runtime falls within one second. This underscores that, in scenarios with adequate computational resources, $\mathcal{SCOPE}$ possesses the potential to offer improved speedup. Consequently, on a robust system equipped with more cores, $\mathcal{SCOPE}$ holds the promise of providing high-quality solutions in less time.

\begin{figure}[ht]
\centering
    \subfigure[\label{fig:budget_score}]{
    \includegraphics[width=0.46\linewidth]{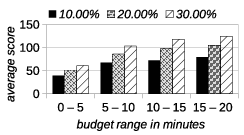}}
    \quad
    \subfigure[\label{fig:budget_time}]{
    \includegraphics[width=0.46\linewidth]{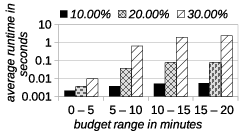}}
    \subfigure[\label{fig:density_score}]{
    \includegraphics[width=0.46\linewidth]{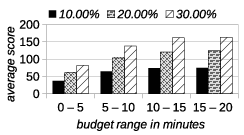}}
    \quad
    \subfigure[\label{fig:density_time}]{
    \includegraphics[width=0.46\linewidth]{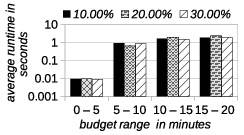}}
\caption{Sensitivity analysis of $\mathcal{SCOPE}$ in Bengaluru road network: (a) \& (b)Average score gain and runtime for different budget overhead ($10\%, 20\%, 30\%$) over the fastest path cost, while $20\%$ of edges with positive score and executed using 24 cores, (c) \& (d)Average score gain and runtime for varying density of edges with positive score ($10\%,20\%, 30\%$), while $30\%$ overhead over the fastest path cost and executed using 24 cores. For (b) \& (d)Y-axis is in $log_{10}$ scale.}
\label{fig:sensitivity}
\end{figure}

\subsubsection{Sensitivity Analysis}
\label{sec:sensitivity}
In this section, we investigate the behaviour of $\mathcal{SCOPE}$ under various combinations of variable parameters in the Bengaluru dataset.

Fig. \ref{fig:sensitivity}(a) and \ref{fig:sensitivity}(b) analyze $\mathcal{SCOPE}$ for different budget overhead. Considering the total budget as the sum of the fastest path's travel time and overhead, this experiment explores overhead of $10\%$, $20\%$, and $30\%$ over the fastest path's travel time. We have created distinct query sets for each overhead value for this experiment. $\mathcal{SCOPE}$ exhibits consistent trends in both average score and average runtime as the overhead increases across each query set. The maximum runtime observed is $2.37$ seconds for query set-4 with a $30\%$ overhead. Considering the linear scalability of $\mathcal{SCOPE}$, these results can be further improved in a system with a higher core count.

In Fig. \ref{fig:sensitivity}(c) and \ref{fig:sensitivity}(d), we evaluate the performance of $\mathcal{SCOPE}$ in networks featuring varying densities of edges with positive scores. For this experiment, we first simulate the Bengaluru dataset with different percentages of edges with a positive score ($10\%, 20\%, 30\%$) to create three distinct datasets. Subsequently, four query sets, as detailed in Section \ref{sec:query-set}, are generated for each dataset. Here, one can observe that the average score achieved by $\mathcal{SCOPE}$ increases as the density of edges with positive score values rises while the runtime remains consistent. This phenomenon occurs because a higher density of edges with positive scores results in more such edges being included in the final solution, consequently leading to a higher score gain. However, $\mathcal{SCOPE}$ persists in exploring the same search space for high-quality solutions, maintaining a consistent runtime.

\section{Conclusion}
\label{sec:conclusion}
We introduced an efficient solution for a novel problem known as the Time-Dependent Constrained Path Optimization problem (TD-CPO). TD-CPO seeks to determine a source-destination path that maximizes its score while adhering to a predefined travel time budget. The solution presented in this paper holds significant potential in urban navigation. Despite having exponential time complexity, the proposed solution $\mathcal{SCOPE}$ produces a superior quality solution for the TD-CPO problem within a reasonable execution time. The efficiency is achieved through adept utilization of temporal properties of the road network. Experimental evaluations on large road networks demonstrate that $\mathcal{SCOPE}$ produces significantly higher quality solutions (with comparable execution times) than the state-of-the-art algorithms. The linear scalability of $\mathcal{SCOPE}$ underscores its applicability in problems with larger search spaces, mainly when supported by higher system configurations.

\section*{Acknowledgements}

This research work is supported by the Indian Institute of Technology Ropar and the University of Hull.





\bibliographystyle{IEEEtran}
\bibliography{sn-bibliography}

\begin{thebibliography}{10}
\providecommand{\url}[1]{#1}
\csname url@samestyle\endcsname
\providecommand{\newblock}{\relax}
\providecommand{\bibinfo}[2]{#2}
\providecommand{\BIBentrySTDinterwordspacing}{\spaceskip=0pt\relax}
\providecommand{\BIBentryALTinterwordstretchfactor}{4}
\providecommand{\BIBentryALTinterwordspacing}{\spaceskip=\fontdimen2\font plus
\BIBentryALTinterwordstretchfactor\fontdimen3\font minus \fontdimen4\font\relax}
\providecommand{\BIBforeignlanguage}[2]{{%
\expandafter\ifx\csname l@#1\endcsname\relax
\typeout{** WARNING: IEEEtran.bst: No hyphenation pattern has been}%
\typeout{** loaded for the language `#1'. Using the pattern for}%
\typeout{** the default language instead.}%
\else
\language=\csname l@#1\endcsname
\fi
#2}}
\providecommand{\BIBdecl}{\relax}
\BIBdecl

\bibitem{bigdata-shashi}
S.~Shekhar, V.~Gunturi, M.~R. Evans, and K.~Yang, ``Spatial big-data challenges intersecting mobility and cloud computing,'' in \emph{Proceedings of the Eleventh ACM International Workshop on Data Engineering for Wireless and Mobile Access}, ser. MobiDE '12.\hskip 1em plus 0.5em minus 0.4em\relax New York, NY, USA: Association for Computing Machinery, 2012, p. 1–6.

\bibitem{csp-foresthop}
Z.~Liu, L.~Li, M.~Zhang, W.~Hua, P.~Chao, and X.~Zhou, ``Efficient constrained shortest path query answering with forest hop labeling,'' in \emph{2021 IEEE 37th International Conference on Data Engineering (ICDE)}, pp. 1763--1774.

\bibitem{shahabi-ils}
Y.~Lu and C.~Shahabi, ``An arc orienteering algorithm to find the most scenic path on a large-scale road network,'' in \emph{Proceedings of the 23rd SIGSPATIAL International Conference on Advances in Geographic Information Systems}, 2015.

\bibitem{kkd-wise22}
K.~K. Dutta, A.~Dewan, and V.~M.~V. Gunturi, ``A multi-threading algorithm for constrained path optimization problem on road networks,'' in \emph{Web Information Systems Engineering -- WISE 2022}, 2022, pp. 110--118.

\bibitem{csp-td-vldb19}
Y.~Yuan and et~al, ``Constrained shortest path query in a large time-dependent graph,'' \emph{Proc. VLDB Endow.}, vol.~12, no.~10, p. 1058–1070, 2019.

\bibitem{csp-td-icde19}
Y.~Yuan, X.~Lian, G.~Wang, L.~Chen, Y.~Ma, and Y.~Wang, ``Weight-constrained route planning over time-dependent graphs,'' in \emph{IEEE 35th ICDE}, 2019, pp. 914--925.

\bibitem{shahabi-scenic}
Y.~Lu, G.~Joss{\'e}, T.~Emrich, U.~Demiryurek, M.~Renz, C.~Shahabi, and M.~Schubert, ``Scenic routes now: Efficiently solving the time-dependent arc orienteering problem,'' in \emph{Proceedings of the 2017 ACM on Conference on Information and Knowledge Management}, pp. 487--496.

\bibitem{kaur-mdm}
R.~Kaur, V.~Goyal, V.~M.~V. Gunturi, A.~Saini, K.~Sanadhya, R.~Gupta, and S.~Ratra, ``A navigation system for safe routing,'' in \emph{22nd {IEEE} International Conference on Mobile Data Management, {MDM} June 15-18, 2021}, 2021, pp. 240--243.

\bibitem{aop-GAVALAS2015}
D.~Gavalas, C.~Konstantopoulos, K.~Mastakas, G.~Pantziou, and N.~Vathis, ``Approximation algorithms for the arc orienteering problem,'' \emph{Information Processing Letters}, vol. 115, no.~2, pp. 313--315, 2015.

\bibitem{kaur-mnpj}
R.~Kaur, V.~Goyal, and V.~M.~V. Gunturi, ``Finding the most navigable path in road networks,'' \emph{GeoInformatica}, vol.~25, no.~1, pp. 207--240, 2021.

\bibitem{tdcrp-icde19}
Y.~Yuan, X.~Lian, G.~Wang, L.~Chen, Y.~Ma, and Y.~Wang, ``Weight-constrained route planning over time-dependent graphs,'' in \emph{2019 IEEE 35th International Conference on Data Engineering (ICDE)}, 2019, pp. 914--925.

\bibitem{GunturiS17}
V.~M.~V. Gunturi and S.~Shekhar, \emph{Spatio-Temporal Graph Data Analytics}, 2017.

\bibitem{GunturiSY15}
V.~M.~V. Gunturi, S.~Shekhar, and K.~Yang, ``A critical-time-point approach to all-departure-time lagrangian shortest paths,'' \emph{{IEEE} Trans. Knowl. Data Eng.}, vol.~27, no.~10, pp. 2591--2603, 2015.

\bibitem{memetic-j}
C.~Chen, L.~Gao, X.~Xie, and Z.~Wang, ``Enjoy the most beautiful scene now: a memetic algorithm to solve two-fold time-dependent arc orienteering problem,'' \emph{Frontiers of Computer Science}, vol.~14, no.~2, pp. 364--377, aug 2019.

\bibitem{memetic-conf}
L.~Gao, C.~Chen, H.~Huang, and C.~Xiang, ``A memetic algorithm for finding the two-fold time-dependent most beautiful driving routes,'' in \emph{2019 IEEE Wireless Communications and Networking Conference (WCNC)}, pp. 1--6.

\bibitem{csp-cola-gpu}
S.~Lu, B.~He, Y.~Li, and H.~Fu, ``Accelerating exact constrained shortest paths on gpus,'' \emph{Proc. VLDB Endow.}, vol.~14, no.~4, p. 547–559, dec 2020.

\bibitem{labeling}
P.~Hansen, ``Bicriterion path problems,'' in \emph{Multiple Criteria Decision Making Theory and Application}.\hskip 1em plus 0.5em minus 0.4em\relax Springer Berlin Heidelberg, 1980, pp. 109--127.

\bibitem{longestpath}
R.~Bulterman and et~al, ``On computing a longest path in a tree,'' \emph{Information Processing Letters}, vol.~81, no.~2, pp. 93--96, 2002.

\bibitem{cyclic-to-dag}
R.~P. Stanley, ``Acyclic orientations of graphs,'' \emph{Discrete Mathematics}, vol.~5, no.~2, pp. 171--178, 1973.

\bibitem{arxiv}
K.~K. Dutta, A.~Dewan, and V.~M.~V. Gunturi, ``A multi-threading algorithm for constrained path optimization problem on road networks,'' \emph{CoRR}, vol. abs/2208.02296, 2022.

\bibitem{datacite}
A.~Karduni and et~al., ``A protocol to convert spatial polyline data to network formats and applications to world urban road networks,'' \emph{Scientific Data}, vol.~3, 2016.

\bibitem{datacite2}
https://users.cs.utah.edu/~lifeifei/SpatialDataset.htm.

\end{thebibliography}

\end{document}